\documentclass[fleqn,twoside]{article}
\usepackage[headings]{espcrc2}

\usepackage{graphicx}

\newcommand{\AmS}{{\protect\the\textfont2
  A\kern-.1667em\lower.5ex\hbox{M}\kern-.125emS}}

\title{Sophisticated momenta mapping with DIANA}

\author{
   M.~Tentyukov\address{Institut f\"ur Theoretische
                              Teilchenphysik, Universit\"at Karlsruhe, Germany
               }\address[BU]{Fakult\"at f\"ur Physik, Universit\"at Bielefeld, Germany
               }\thanks{On leave from BLTP, Joint Institute
                        for Nuclear Research, Dubna, Russia
               }
   and J.~Fleischer\addressmark[BU]
       }

\begin{document}

\begin{abstract}
Details of recent developments of the
Feynman diagram analyzer  DIANA (DIagram ANAlyser) are presented. 
Apart from some discussion
about QGRAF and new plotting features, we concentrate on a new
sophisticated mechanism of momenta mapping.
\vspace{1pc}
\end{abstract}

\maketitle

The C-program DIANA (DIagram ANAlyser) for the automatic Feynman diagram
evaluation was first documented in \cite{Diana}\footnote[1]{See also\\
http://www.physik.uni-bielefeld.de/$\tilde{\,}$tentukov/diana.html}.
Meanwhile there exists a growing series of applications \cite{usage}
for which DIANA was used.

The current DIANA version is 2.37. Here we describe some details of 
the recent development.

\section{Compatibility with current QGRAF versions}

DIANA uses QGRAF \cite{QGRAF} as diagram generator. By now there exist several
QGRAF versions which, however, have incompatible syntax of their
input files. In most cases the user does not need or does not want to
know which version of QGRAF is in use. DIANA takes care of this. It
automatically investigates which version of QGRAF is implemented.
Starting from version 2.37 it also provides the 
user with information about the version used for the diagram 
generation.
%Therefore DIANA was made more flexible to realize which version
%is implemented and even, starting from version 2.37, provides the 
%user with information about the version used for the diagram 
%generation.
At present, all QGRAF versions $2. \cdots$ and $3. \cdots$ are supported.
From the DIANA point of view there is no difference between them; the
differences concern only syntax.

\section{New plotting features}
QGRAF cannot generate two-line vertices. This causes problems
when  DIANA is requested to introduce counterterms. In this
case it produces three-line
%In case DIANA is requested to introduce counterterms, for the reason
%that QGRAF cannot generate two-line vertices, DIANA produces three-line
vertices but does not show one of them, i.e. in this case DIANA skips
the image of the ``spurious'' line in the diagram. Another new feature
is introduced to suppress labels. In version 2.35 of DIANA and later 
ones these options (changing the model extensions \cite{Chme})
are taken into account in terms of ``optional
extensions'' in the propagators of the model file, e.g. in
\begin{verbatim}
[le,Le;l; FF(num,fnum,vec, mle)*i_;
                 mmle; nothing,0,0;]
\end{verbatim}
the last two extensions \verb|;nothing,0,0;]| (separated by ;) have the 
following meaning:\\
1. the new line type of the propagator ``nothing'' means that this line
will not be drawn. The two following parameters in general stand
for line characterizations and must be included in any case.\\
2. the second ``extension'' apparently is empty, which means that no
label will be drawn - in previous versions this meant the ``default image''
(in the above example: \verb|le|)

\section{Automation of momenta distribution}
Again, starting from version 2.25, momenta can be introduced automatically.
There are several different approaches \cite{acat02,acat03}.

As  ``standard'' way of attributing momenta, we consider the possibility
to calculate, e.g., from external momenta the so called ``chords'', characterizing
the virtual lines apart from the integration momenta.

A more specific problem is the following: occasionally topologies 
have to be 
generated from more complicated ones by scratching lines and one may
want to stick to the momenta introduced for the lines which are kept. 
This option was described in \cite{acat02}.

A new and more difficult situation occurs when the user wants to ``scratch'' some
external lines. Consider e.g. the Anomalous Magnetic Moment; this is a 3-point
function of the type
 fermion -- fermion with an external photon of zero momentum. 
Performing a necessary differentiation
 and putting the photon momentum to zero, the
resulting graph appears to be a 2-point function with higher powers of
scalar propagators. In this case we need to map the momenta of the 3-point 
function to the
corresponding momenta of a 2-point function, putting one external momentum to
zero.

At present, DIANA is able to map momenta from  $n$-point
functions to m-point functions ($m<n$). Here we provide some details.

\begin{figure}[ht]
\begin{center}
\includegraphics[width=\linewidth]{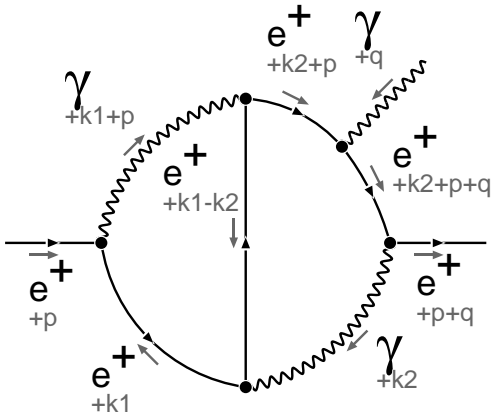}
\end{center}
\caption{
\label{fig1}
A typical diagram for the two-loop Anomalous Magnetic Moment.
Arrows indicate the direction of momenta.
}
\end{figure}

Let us consider one of the diagrams relevant for  $g-2$ in two loop
approximation, 
Fig.\ref{fig1}.

After differentiation w.r.t. to the photon momentum $q$, we have to put $q=0$ 
and evaluate the resulting propagator with the only one momentum $p$.

\begin{figure}[ht]
\begin{center}
\includegraphics[width=\linewidth]{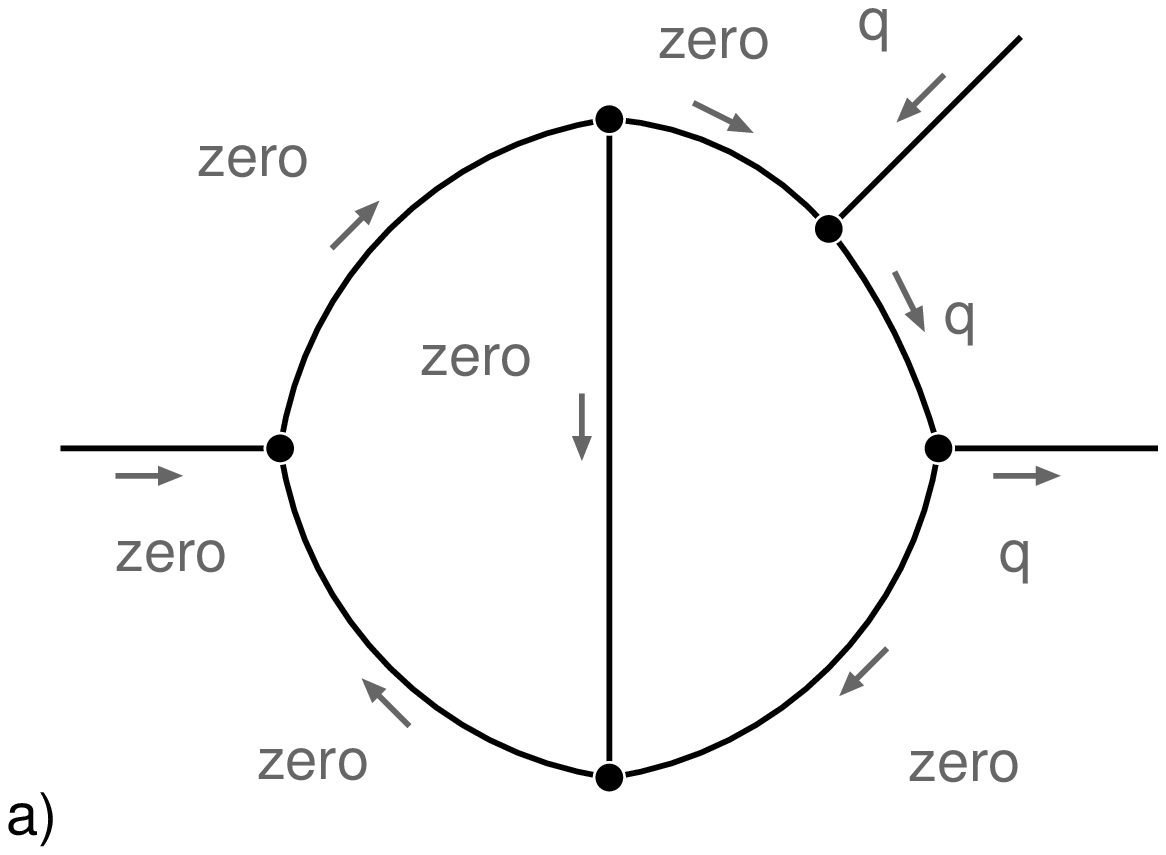}
\end{center}
\begin{center}
\includegraphics[width=\linewidth]{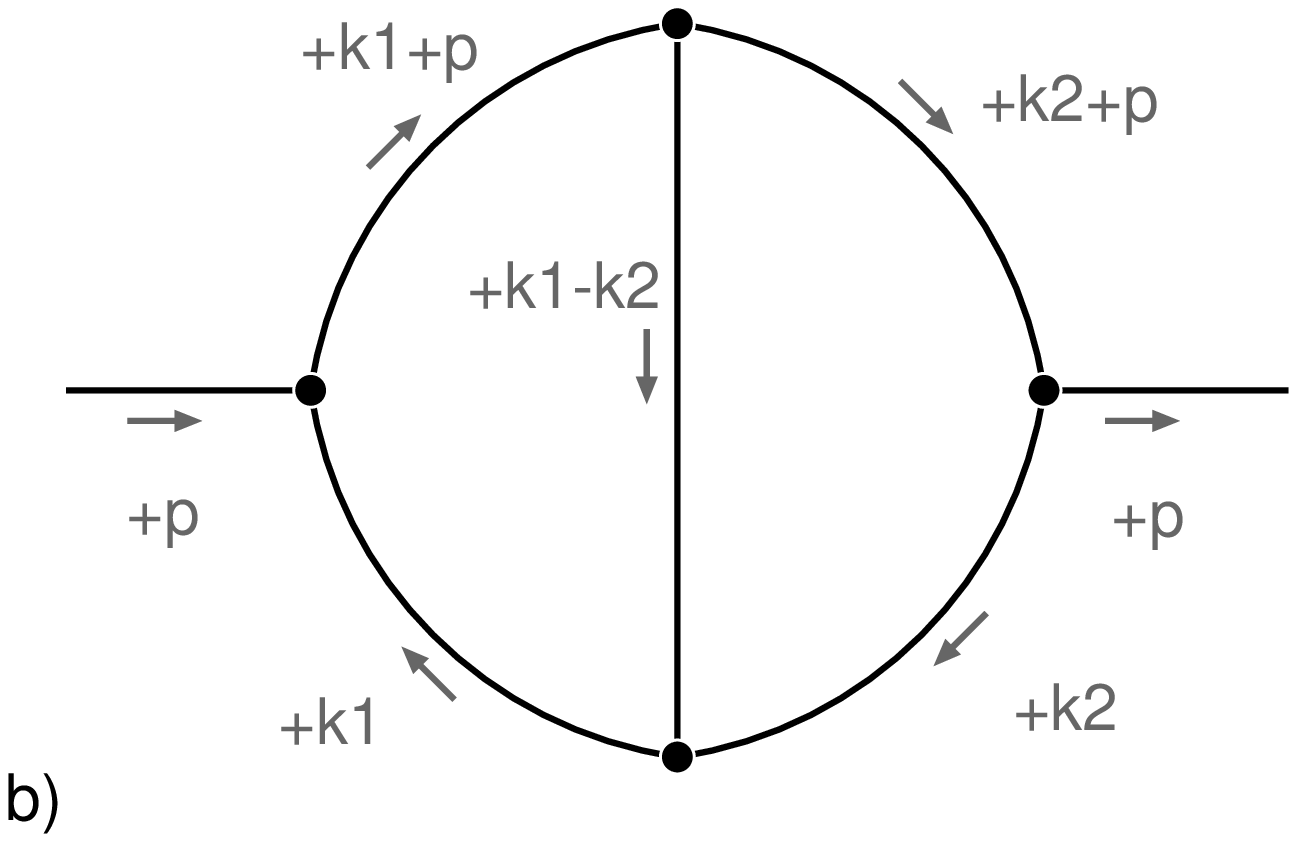}
\end{center}
\caption{
\label{fig2}
Momenta decomposition. The initial momentum flow  is a ``sum'' of a
three-point momentum flow  a) and a two-point momentum flow b).
}
\end{figure}

To perform the differentiation w.r.t. $q$ we formally consider
at first a
diagram with the only nonvanishing momentum $q$, all others zero. In a FORM
program the differentiation is then performed w.r.t. $q$ and after that for 
zero momentum $q$, momenta fitting to the three-point function are introduced,
see Fig.\ref{fig2}.

In a first step we generate the FORM input with momenta as in Fig.\ref{fig2}a).
After differentiation w.r.t. $q$, we look (e.g. in another program 
like  MINCER \cite{MINCER}) for
the corresponding two-point topology (Fig.\ref{fig2}b),
add for technical reasons all momenta from the two-point topology \ref{fig2}b) to the 
current three-point topology \ref{fig2}a), put $q=0$ and generate the FORM 
input for further evaluation.

\begin{figure}[ht]
\begin{center}
\includegraphics[width=\linewidth]{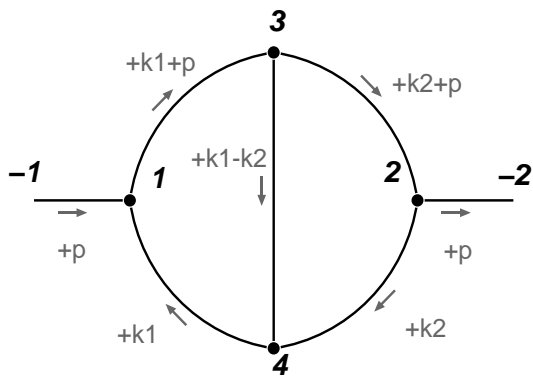}
\end{center}
\caption{
\label{fig3}
Topology of \protect\ref{fig2}b) as a set of ordered pairs of numbers
like (fromvertex, tovertex).
}
\end{figure}

The main problem here is to determine which two-point topology corresponds to
the current three-point topology. In DIANA 
topologies are represented \cite{Diana} in terms of
    ordered pairs of numbers like (fromvertex, tovertex) (see
Fig.~\ref{fig3}).

All external legs have negative numbers. The
momenta distribution according to Fig. \ref{fig3}, e.g., is added like
 {
\small\begin{verbatim}
topology A = 
  (-2,2)(-1,1)(1,3)(4,1)(2,4)(3,2)(3,4):
               k1+p,k1,k2,k2+p,k1-k2;
\end{verbatim}
}
This fixes directions and values of all momenta on the internal lines.
External lines must be known from the process definition.

A special TM\footnote{TM language is used
  by DIANA, see \cite{Diana}} operator was introduced:
\begin{verbatim}
\amputateTopology(varname, tablename)
\end{verbatim}
It tries to find an amputated topology 
for the diagram under consideration (e.g. \ref{fig2}a, taking off $q$)
in the table ``tablename'' (e.g. from MINCER) and
returns this topology (e.g. \ref{fig2}b).
It adds all momenta from the table topology to momenta of the current (\ref{fig2}a)
topology and saves the resulting momenta to the local variable \verb|<varname>_M|.

On failure this TM operator returns an 
empty string. If it can't load the table it halts the program. 

Apart from \verb|<varname>_M|, the operator creates several other variables:
\begin{itemize}
\item
\verb|<varname>_IN| - number of internal lines in the amputated topology;
\item
\verb|<varname>_N| - name of the found topology.

\end{itemize}

The following arrays {\tt amputated(original)} are indices of lines of the amputated
topology which corresponds to the original:
\begin{itemize}
\item
\verb|<varname>_E| - external lines of original \verb|<->| external lines of
amputated; 
\item
\verb|<varname>_IE| internal lines of original \verb|<->| internal
lines of amputated; 
\item
\verb|<varname>_I| internal lines of original \verb|<->|
external lines of amputated.
\end{itemize}
The absolute value of elements of these arrays are indices of the lines and their
signs correspond to their direction.

External lines to be removed by the operator \verb|\amputateTopology|
are marked by another TM operator, 
\begin{verbatim}
\rmExtLeg(n)
\end{verbatim}
where \verb|n| is the index of the external line. There can be more than
only one external line which the user wants to be removed.

\end{document}